\documentclass[aps,prb,twocolumn,showpacs]{revtex4}

\usepackage{graphicx}
\usepackage{amsmath}
\usepackage{amssymb}

\begin{document}

\title{Pseudogap and charge dynamics in doped cuprates}

\author{Ling Qin}

\affiliation{Department of Physics, Beijing Normal University, Beijing 100875, China}

\author{Jihong Qin}

\affiliation{Department of Physics, University of Science and Technology Beijing, Beijing 100083, China}

\author{Shiping Feng}

\email{spfeng@bnu.edu.cn}

\affiliation{Department of Physics, Beijing Normal University, Beijing 100875, China ~~~}

\begin{abstract}
Within the microscopic theory of the normal-state pseudogap state, the doping and temperature dependence of the charge
dynamics in doped cuprates is studied in the whole doping range from the underdoped to heavily overdoped. The
conductivity spectrum in the underdoped and optimally doped regimes contains the low-energy non-Drude peak and unusual
midinfrared band. However, the position of the midinfrared band shifts towards to the low-energy non-Drude peak with
increasing doping. In particular, the low-energy non-Drude peak incorporates with the midinfrared band in the heavily
overdoped regime, and then the low-energy Drude behavior recovers. It is shown that the striking behavior of the
low-energy non-Drude peak and unusual midinfrared band in the underdoped and optimally doped regimes is closely related
to the emergence of the doping and temperature dependence of the normal-state pseudogap.
\end{abstract}


\maketitle

\section{Introduction}

Since the discovery of superconductivity in doped cuprtes, a significant body of reliable and reproducible data has
been accumulated by using many probes
\cite{Tannern92,Cooper94,Kastner98,Timusk99,Puchkov96,Cooper09,Eschrig06,Devereaux07,Hufner08,Chatterjeea11}, which
show that the most remarkable expression of the nonconventional physics is found in the normal-state. The normal-state
properties in the underdoped regime exhibit a number of anomalous properties in sense that they do not fit in the
conventional Fermi-liquid theory. In particular, it is widely believed that the anomalous normal-state properties in
the underdoped regime are closely related to a normal-state pseudogap
\cite{Kastner98,Timusk99,Puchkov96,Cooper09,Eschrig06,Devereaux07,Hufner08,Chatterjeea11}, since this normal-state
pseudogap observed in the excitation spectrum as a suppression of the spectral weight is particular large in the
underdoped regime, then it smoothly decreases with increasing doping \cite{Hufner08,Chatterjeea11}.

After intensive investigations over more than two decades, many ideas have been proposed to understand the origin of
the ubiquitous normal-state pseudogap and its connection to the anomalous normal-state properties. Because early
experiments indicate a dominated d-wave normal-state pseudogap compatible with the superconducting gap, some authors
argued that the normal-state pseudogap is related to some form of the preformed pairing \cite{Norman98}. On the other
hand, there are many reports suggesting that the normal-state pseudogap crossover temperature is associated with a
broken symmetry \cite{Daou10}, and thus another competing order parameter, such as a density wave order
\cite{Chakravarty01}. Moreover, a phenomenological theory of the normal-state pseudogap state has been developed
\cite{Yang06}, where a new feature is the presence of an additional energy scale, i.e., the normal-state pseudogap in a
doped resonant valence bond state. Furthermore, it has been argued that the pseudogap is a combination of a quantum
disordered d-wave superconductor and an entirely different form of competing order, originating from the particle-hole
channel \cite{Tesanovic08}. Recently, the interplay between the normal-state pseudogap state and superconductivity in
doped cuprates has been discussed based on the kinetic energy driven superconducting mechanism \cite{feng12}, where the
charge carriers interaction directly through the kinetic energy by exchanging spin excitations induces the normal-state
pseudogap state in the particle-hole channel and superconducting-state in the particle-particle channel, therefore both
the normal-state pseudogap and superconducting gap are dominated by one energy scale, and they are the result of the
strong electron correlation. In particular, this normal-state pseudogap is closely related to the quasiparticle
coherent weight, and therefore it suppresses the spectral weight. This microscopic normal-state pseudogap theory gives
a consistent description of the physical properties of doped cuprates in the pseudogap phase \cite{Zhao12,Zhao121},
including the humplike anomaly of the specific-heat and the unusual evolution of the Fermi arc length with doping and
temperature. In particular, it has been shown within this theoretical framework that the particle-hole asymmetry
electronic state in doped cuprates is a natural consequence due to the presence of the normal-state pseudogap
\cite{Zhao121}.

Among the striking features of the normal-state properties in doped cuprates in the underdoped regime, the physical
quantity which most evidently displays the signature for the normal-state pseudogap is the charge transport
\cite{Tannern92,Cooper94,Kastner98,Timusk99,Puchkov96,Cooper09}, which is manifested by the optical conductivity and
resistivity. The optical studies of the electron excitations have revealed much about the nature of the charge carriers
in doped cuprates. In particular, the normal-state pseudogap can be observed directly by the infrared measurements of
the optical conductivity. Experimentally, it has been shown that the charge dynamics is rather universal within the
whole cuprates
\cite{Tannern92,Cooper94,Kastner98,Timusk99,Puchkov96,Cooper09,Takagi92,Batlogg94,Ando01,Orenstein90,Uchida91,Puchkov96a,Basov96,Homes04},
where the optical conductivity for the same doping is nearly materials independent both in the magnitude and energy
dependence, and in the underdoped regime it shows the low-energy non-Drude behavior (the conductivity decays as
$\rightarrow 1/\omega$ at low energies) and unusual midinfrared band at higher energies. Although the optical
conductivity of doped cuprates is well-established by now
\cite{Tannern92,Cooper94,Kastner98,Timusk99,Puchkov96,Cooper09,Takagi92,Batlogg94,Ando01,Orenstein90,Uchida91,Puchkov96a,Basov96,Homes04},
its full understanding is still a challenging issue. In our early studies \cite{feng04,feng97}, the charge transport
of doped cuprates in the underdoped regime has been discussed by considering the second-order correction for the charge
carriers, and the results are qualitatively consistent with the corresponding experimental data. In this paper, as a
complement of our previous analysis of the charge transport in doped cuprates, we start from the microscopic theory of
the normal-state pseudogap state \cite{feng12} to discuss the doping and temperature dependence of the normal-state
optical properties in the whole doping range from the underdoped to heavily overdoped, and qualitatively reproduce all
main features of the optical measurements on doped cuprates
\cite{Tannern92,Cooper94,Kastner98,Timusk99,Puchkov96,Orenstein90,Uchida91,Puchkov96a,Basov96,Homes04}. In particular,
we show that the anomalous behavior of the optical conductivity in the underdoped regime can be attributed to the
emergence of the normal-state pseudogap.

The paper is organized as follows. The basic formalism is presented in Sec. \ref{framework}, where we generalize the
microscopic theory \cite{feng12} of the normal-state pseudogap state from the form for the discussion of the interplay
between the normal-state pseudogap state and superconductivity to the form for discussions of the normal-state
properties, and then evaluate explicitly the optical conductivity. Within this theoretical framework, we discuss the
charge dynamics in doped cuprates in Sec. \ref{transport}, and then provide a natural explanation to the unusual
conductivity spectrum. Finally, we give a summary in Sec. \ref{conclusions}.

\section{Theoretical framework}\label{framework}

In doped cuprates, the single common feature is the presence of the two-dimensional CuO$_{2}$ plane, and it is believed
that the unconventional physical properties of doped cuprates are closely related to the doped CuO$_{2}$ planes
\cite{Kastner98,Timusk99}. In this case, it has been argued that the $t$-$J$ model on a square lattice captures the
essential physics of the doped CuO$_{2}$ plane \cite{anderson87},
\begin{eqnarray}\label{tjham}
H&=&-t\sum_{l\hat{\eta}\sigma}C^{\dagger}_{l\sigma}C_{l+\hat{\eta}\sigma}+t'\sum_{l\hat{\tau}\sigma}
C^{\dagger}_{l\sigma}C_{l+\hat{\tau}\sigma}\nonumber\\
&+&\mu\sum_{l\sigma}C^{\dagger}_{l\sigma}C_{l\sigma}+J\sum_{l\hat{\eta}}
{\bf S}_{l}\cdot {\bf S}_{l+\hat{\eta}},
\end{eqnarray}
where the summation is over all sites $l$, and for each $l$, over its nearest-neighbors (NN) $\hat{\eta}$ or the next
NN $\hat{\tau}$, ${\bf S}_{l}=(S^{\rm x}_{l},S^{\rm y}_{l},S^{\rm z}_{l})$ are spin operators, and $\mu$ is the
chemical potential. The $t$-$J$ model (\ref{tjham}) is subject to an important on-site local constraint to avoid the
double occupancy, i.e., $\sum_{\sigma}C^{\dagger}_{l\sigma}C_{l\sigma}\leq 1$. In this $t$-$J$ model (\ref{tjham}),
the strong electron correlation manifests itself by this single occupancy local constraint, and therefore the crucial
requirement is to impose this local constraint. It has been shown that this constraint can be treated properly in
analytical calculations within the charge-spin separation (CSS) fermion-spin theory \cite{feng04}, where the physics
of no double occupancy is taken into account by representing the electron as a composite object created by
$C_{l\uparrow}= h^{\dagger}_{l\uparrow}S^{-}_{l}$ and $C_{l\downarrow}=h^{\dagger}_{l\downarrow}S^{+}_{l}$, with the
spinful fermion operator $h_{l\sigma}=e^{-i\Phi_{l\sigma}}h_{l}$ that describes the charge degree of freedom of the
electron together with some effects of spin configuration rearrangements due to the presence of the doped hole itself
(charge carrier), while the spin operator $S_{l}$ represents the spin degree of freedom of the electron, then the
electron single occupancy local constraint is satisfied in analytical calculations. In particular, it has been shown
that under the decoupling scheme, this CSS fermion-spin representation is a natural representation of the constrained
electron defined in the Hilbert subspace without double electron occupancy \cite{feng08}. Furthermore, these charge
carrier and spin are gauge invariant \cite{feng04,feng08}, and in this sense they are real and can be interpreted as
physical excitations \cite{laughlin97}. In this CSS fermion-spin representation, the $t$-$J$ model (\ref{tjham}) can
be expressed as \cite{feng04,feng08},
\begin{eqnarray}\label{cssham}
H&=&t\sum_{l\hat{\eta}}(h^{\dagger}_{l+\hat{\eta}\uparrow}h_{l\uparrow}S^{+}_{l}S^{-}_{l+\hat{\eta}}+
h^{\dagger}_{l+\hat{\eta}\downarrow}h_{l\downarrow}S^{-}_{l}S^{+}_{l+\hat{\eta}})\nonumber\\
&-&t'\sum_{l\hat{\tau}}
(h^{\dagger}_{l+\hat{\tau}\uparrow}h_{l\uparrow}S^{+}_{l}S^{-}_{l+\hat{\tau}}+h^{\dagger}_{l+\hat{\tau}\downarrow}
h_{l\downarrow}S^{-}_{l}S^{+}_{l+\hat{\tau}})\nonumber\\
&-&\mu\sum_{l\sigma}h^{\dagger}_{l\sigma}h_{l\sigma}+J_{{\rm eff}}\sum_{l\hat{\eta}}{\bf S}_{l}\cdot
{\bf S}_{l+\hat{\eta}},
\end{eqnarray}
with $J_{{\rm eff}}=(1-\delta)^{2}J$, and
$\delta=\langle h^{\dagger}_{l\sigma}h_{l\sigma}\rangle=\langle h^{\dagger}_{l}h_{l}\rangle$ is the charge carrier
doping concentration. As an important consequence, the kinetic energy term in the $t$-$J$ model has been transferred
as the interaction between charge carriers and spins, which reflects that even the kinetic energy term in the $t$-$J$
model has a strong Coulombic contribution due to the restriction of no double occupancy of a given site.

The interaction between charge carriers and spins in the $t$-$J$ model (\ref{cssham}) is quite strong, and therefore it
dominates the essential physics in doped cuprates in the doped regime without an antiferromagnetic long-range order.
For the qualitative comparison with the experimental results of doped cuprates, the charge dynamics should be treated
by considering the charge carrier and spin fluctuations. In this case, we follow the previous discussions \cite{guo06},
and obtain the full charge carrier Green's function as,
\begin{eqnarray}\label{Green's-function-1}
g({\bf k},\omega)={1\over\omega-\xi_{\bf k}-\Sigma^{({\rm h})}({\bf k},\omega)},
\end{eqnarray}
where the mean-field (MF) charge carrier spectrum
$\xi_{\bf k}=Zt\chi_{1}\gamma_{{\bf k}}-Zt'\chi_{2}\gamma_{{\bf k}}'-\mu$, the spin correlation functions
$\chi_{1}=\langle S^{+}_{l}S^{-}_{l+\hat{\eta}}\rangle$ and $\chi_{2}=\langle S_{l}^{+}S_{l+\hat{\tau}}^{-}\rangle$,
$\gamma_{{\bf k}}=(1/Z)\sum_{\hat{\eta}}e^{i{\bf k}\cdot\hat{\eta}}$,
$\gamma_{{\bf k}}'=(1/Z)\sum_{\hat{\tau}}e^{i{\bf k}\cdot\hat{\tau}}$, and $Z$ is the number of the NN or next NN
sites, while the self-energy $\Sigma^{({\rm h})}({\bf k},\omega)$ can be evaluated in terms of the spin bubble as
\cite{guo06},
\begin{eqnarray}\label{self-energy-1}
\Sigma^{({\rm }h)}({\bf k},i\omega_{n})&=&{1\over N^{2}}\sum_{{\bf p,p'}}\Lambda^{2}_{{\bf p}+{\bf p}'+{\bf k}}\nonumber\\
&\times&{1\over \beta}\sum_{ip_{m}}g({\bf p}+{\bf k},ip_{m}+i\omega_{n})\Phi({\bf p},{\bf p}',ip_{m}),~~~~
\end{eqnarray}
with $\Lambda_{{\bf k}}=Zt\gamma_{\bf k}-Zt'\gamma_{\bf k}'$, and the spin bubble,
\begin{eqnarray}\label{spin-bubble}
\Phi({\bf p},{\bf p}',ip_{m})={1\over\beta}\sum_{ip'_{m}}D({\bf p}',ip_{m}')D({\bf p}'+{\bf p},ip_{m}'+ip_{m}).~~~
\end{eqnarray}
In the following discussions, we limit the spin part to the MF level, since the electronic structure of doped cuprates
in the normal-state can be well described at this level \cite{guo06}. In this case, the full spin Green's function,
$D({\bf p},\omega)$ in the spin bubble (\ref{spin-bubble}) can be replaced by the MF spin Green's function,
$D^{(0)}({\bf p},\omega)=B_{\bf p}/[\omega^{2}-\omega_{\bf p}^{2}]$, with the MF spin excitation spectrum
$\omega_{\bf p}$ and $B_{\bf p}$ have been given in Ref. \cite{guo06}. This self-energy
$\Sigma^{({\rm h})}({\bf k},\omega)$ renormalizes the MF charge carrier spectrum \cite{guo06}, and therefore it
describes the charge carrier quasiparticle coherence. Moreover, $\Sigma^{({\rm h})}({\bf k},\omega)$ can be broken
up into its symmetric and antisymmetric parts as, $\Sigma^{({\rm h})}({\bf k},\omega)=
\Sigma^{({\rm h})}_{\rm e}({\bf k},\omega)+\omega\Sigma^{({\rm h})}_{\rm o}({\bf k},\omega)$, and then both
$\Sigma^{({\rm h})}_{\rm e}({\bf k},\omega)$ and $\Sigma^{({\rm h})}_{\rm o}({\bf k},\omega)$ are an even function of
$\omega$. Furthermore, the antisymmetric part $\Sigma^{({\rm h})}_{\rm o}({\bf k},\omega)$ is closely related with the
charge carrier quasiparticle coherent weight as
$Z^{-1}_{\rm hF}({\bf k},\omega)=1-{\rm Re}\Sigma^{({\rm h})}_{\rm o}({\bf k},\omega)$, and therefore it reduces the
charge carrier quasiparticle bandwidth, and then the energy scale of the charge carrier quasiparticle band is
controlled by the magnetic interaction $J$, while the symmetric part
${\rm Re}\Sigma^{({\rm h})}_{\rm e}({\bf k},\omega)$ may be a constant, independent of (${\bf k},\omega$), i.e., it
just renormalizes the chemical potential, and therefore can be neglected. In this case, in the static limit
approximation for the quasiparticle coherent weight,
i.e., $Z^{-1}_{\rm hF}=1-{\rm Re}\Sigma^{({\rm h})}_{\rm o}({\bf k},\omega=0)|_{{\bf k}=[\pi,0]}$, the full charge
carrier Green's function has been obtained as \cite{guo06},
\begin{eqnarray}\label{Green's-function-2}
g({\bf k},\omega)&=&{Z_{\rm hF}\over \omega-\bar{\xi_{{\bf k}}}},
\end{eqnarray}
where the renormalized charge carrier quasiparticle spectrum $\bar{\xi}_{{\bf k}}=Z_{\rm hF}\xi_{{\bf k}}$. With the
help of this charge carrier Green's function (\ref{Green's-function-2}), the self-energy
$\Sigma^{({\rm h})}({\bf k},\omega)$ in Eq. (\ref{self-energy-1}) has been evaluated explicitly as \cite{guo06},
\begin{widetext}
\begin{eqnarray}\label{self-energy-2}
\Sigma^{({\rm }h)}({\bf k},\omega)&=&{1\over N^{2}}\sum_{{\bf p}{\bf p'}}Z_{\rm hF}
\Lambda^{2}_{{\bf p}+{\bf p'}+{\bf k}}{B_{\bf p'}B_{{\bf p}+{\bf p'}}\over 4\omega_{\bf p'}\omega_{{\bf p}+{\bf p'}}}
\left ({F_{1}({\bf k},{\bf p},{\bf p'})\over\omega-\omega_{\bf p'}+\omega_{{\bf p}+{\bf p'}}
-\bar{\xi}_{{\bf p}+{\bf k}}}+{F_{2}({\bf k},{\bf p},{\bf p'})\over\omega+\omega_{\bf p'}
-\omega_{{\bf p}+{\bf p'}}-\bar{\xi}_{{\bf p}+{\bf k}}}\right .\nonumber\\
&+&\left .{F_{3}({\bf k},{\bf p},{\bf p'})\over\omega+\omega_{\bf p'}+\omega_{{\bf p}+{\bf p'}}
-\bar{\xi}_{{\bf p}+{\bf k}}}+{F_{4}({\bf k},{\bf p},{\bf p'})\over\omega-\omega_{\bf p'}-\omega_{{\bf p}+{\bf p'}}
-\bar{\xi}_{{\bf p}+{\bf k}}}\right ),
\end{eqnarray}
\end{widetext}
where the kernel functions $F_{1}({\bf k},{\bf p},{\bf p'})=n_{\rm F}(\bar{\xi}_{{\bf p}+{\bf k}})
[n_{B}(\omega_{\bf p'})-n_{\rm B}(\omega_{{\bf p}+{\bf p'}})]-n_{\rm B}(\omega_{{\bf p}+{\bf p'}})
n_{\rm B}(-\omega_{\bf p'})$, $F_{2}({\bf k},{\bf p},{\bf p'})=n_{\rm F}(\bar{\xi}_{{\bf p}+{\bf k}})
[n_{\rm B}(\omega_{{\bf p'}+{\bf p}})-n_{\rm B}(\omega_{\bf p'})]-n_{\rm B}(\omega_{\bf p'})
n_{\rm B}(-\omega_{{\bf p'}+{\bf p}})$, $F_{3}({\bf k},{\bf p},{\bf p'})=n_{\rm F}(\bar{\xi}_{{\bf p}+{\bf k}})
[n_{\rm B}(\omega_{{\bf p}+{\bf p'}})-n_{\rm B}(-\omega_{\bf p'})]+n_{\rm B}(\omega_{\bf p'})
n_{\rm B}(\omega_{{\bf p}+{\bf p'}})$, $F_{4}({\bf k},{\bf p},{\bf p'})=n_{\rm F}(\bar{\xi}_{{\bf p}+{\bf k})}
[n_{\rm B}(-\omega_{\bf p'})-n_{\rm B}(\omega_{{\bf p}+{\bf p'}})]+n_{\rm B}(-\omega_{\bf p'})
n_{\rm B}(-\omega_{{\bf p}+{\bf p'}})$, and $n_{\rm B}(\omega)$ and $n_{\rm F}(\omega)$ are the boson and fermion
distribution functions, respectively. In the above discussions, the charge carrier quasiparticle coherent weight
$Z_{\rm hF}$ and all other order parameters have been determined simultaneously by the self-consistent calculation
\cite{guo06}. In this sense, the above calculations are exact without using adjustable parameters.

However, for a complete description of the normal-state pseudogap state, the self-energy
$\Sigma^{({\rm h})}({\bf k},\omega)$ in Eq. (\ref{self-energy-1}) also can be rewritten approximately as \cite{feng12},
\begin{eqnarray}\label{self-energy-3}
\Sigma^{({\rm }h)}({\bf k},\omega)&\approx&{[2\bar{\Delta}_{\rm pg}({\bf k})]^{2}\over \omega+M_{\bf k}},
\end{eqnarray}
where $M_{\bf k}$ is the energy spectrum of $\Sigma^{({\rm h})}({\bf k},\omega)$. Since the interaction force and
normal-state pseudogap have been incorporated into $\bar{\Delta}_{\rm pg}({\bf k})$, it is called as the effective
normal-state pseudogap. From this expression of the self-energy in Eq. (\ref{self-energy-3}), we therefore find that
the quasiparticle coherent weight
$Z^{-1}_{\rm hF}=1+[2\bar{\Delta}_{\rm pg}({\bf k})]^{2}/M^{2}_{\bf k}|_{{\bf k}=[\pi,0]}$, which reflects that the
main effect of the normal-state pseudogap has been contained in the quasiparticle coherent weight. In the following
discussions, we focus on the discussions of the normal-state pseudogap state beyond above static limit approximation
\cite{guo06} for the self-energy $\Sigma^{({\rm h})}({\bf k},\omega)$, and show explicitly that one quasiparticle band
in the full charge carrier Green's function (\ref{Green's-function-2}) is split into two branches. Substituting the
self-energy $\Sigma^{({\rm h})}({\bf k},\omega)$ in Eq. (\ref{self-energy-3}) into Eq. (\ref{Green's-function-1}), the
full charge carrier Green's function in the presence of the normal-state pseudogap is obtained explicitly as,
\begin{eqnarray}\label{Green's-function-3}
g({\bf k},\omega)={\alpha^{\rm (n)}_{1{\bf k}}\over\omega-E^{+}_{{\rm h}{\bf k}}}+{\alpha^{\rm (n)}_{2{\bf k}}
\over\omega -E^{-}_{{\rm h}{\bf k}}},
\end{eqnarray}
where the charge carrier quasiparticle coherence factors $\alpha^{\rm (n)}_{1{\bf k}}=(E^{+}_{{\rm h}{\bf k}}
+M_{\bf k})/(E^{+}_{{\rm h}{\bf k}}-E^{-}_{{\rm h}{\bf k}})$ and $\alpha^{\rm (n)}_{2{\bf k}}=-(E^{-}_{{\rm h}{\bf k}}
+M_{\bf k})/(E^{+}_{{\rm h}{\bf k}}-E^{-}_{{\rm h}{\bf k}})$ satisfy the sum rule:
$\alpha^{({\bf n})}_{1{\bf k}}+\alpha^{({\bf n})}_{2{\bf k}}=1$, and there are two branches of the charge carrier
quasiparticle spectrum due to the presence of the normal-state pseudogap,
$E^{+}_{{\rm h}{\bf k}}=[\xi_{{\bf k}}-M_{\bf k}+\sqrt{(\xi_{{\bf k}}+M_{\bf k})^{2}
+16\bar{\Delta}^{2}_{\rm pg}({\bf k})}]/2$ and $E^{-}_{{\rm h}{\bf k}}=[\xi_{{\bf k}}-M_{\bf k}-\sqrt{(\xi_{{\bf k}}
+M_{\bf k})^{2}+16\bar{\Delta}^{2}_{\rm pg}({\bf k})}]/2$, while $\bar{\Delta}_{\rm pg}({\bf k})$ and $M_{\bf k}$ can
be obtained directly from the self-energy $\Sigma^{({\rm h})}({\bf k},\omega)$ in Eq. (\ref{self-energy-2}) as,
\begin{subequations}
\begin{eqnarray}
\bar{\Delta}_{\rm pg}({\bf k})&=&{L^{\rm (n)}_{2}({\bf k})\over 2\sqrt{L^{\rm (n)}_{1}({\bf k})}},\label{pseudogap}\\
M_{\bf k}&=&{L^{\rm (n)}_{2}({\bf k})\over L^{\rm (n)}_{1}({\bf k})},
\end{eqnarray}
\end{subequations}
with the functions $L^{\rm (n)}_{1}({\bf k})$ and $L^{\rm (n)}_{2}({\bf k})$ are given by,
\begin{widetext}
\begin{subequations}
\begin{eqnarray}
L^{\rm (n)}_{1}({\bf k})&=&{1\over N^{2}}\sum_{{\bf p}{\bf p'}}Z_{\rm hF}\Lambda^{2}_{{\bf p}+{\bf p'}+{\bf k}}
{B_{\bf p'}B_{{\bf p}+{\bf p'}}\over 4\omega_{\bf p'}\omega_{{\bf p}+{\bf p'}}}\left ({F_{1}({\bf k},{\bf p},{\bf p'})
\over (\omega_{\bf p'}-\omega_{{\bf p}+{\bf p'}}+\bar{\xi}_{{\bf p}+{\bf k}})^{2}}+{F_{2}({\bf k},{\bf p},{\bf p'})
\over (\omega_{\bf p'}-\omega_{{\bf p}+{\bf p'}}-\bar{\xi}_{{\bf p}+{\bf k}})^{2}}\right .\nonumber\\
&+&\left .{F_{3}({\bf k},{\bf p},{\bf p'})\over (\omega_{\bf p'}+\omega_{{\bf p}+{\bf p'}}
-\bar{\xi}_{{\bf p}+{\bf k}})^{2}}+{F_{4}({\bf k},{\bf p},{\bf p'})\over (\omega_{\bf p'}+\omega_{{\bf p}+{\bf p'}}
+\bar{\xi}_{{\bf p}+{\bf k}})^{2}}\right ),\\
L^{\rm (n)}_{2}({\bf k})&=&{1\over N^{2}}\sum_{{\bf p}{\bf p'}}Z_{\rm hF}\Lambda^{2}_{{\bf p}+{\bf p'}+{\bf k}}
{B_{\bf p'}B_{{\bf p}+{\bf p'}}\over 4\omega_{\bf p'}\omega_{{\bf p}+{\bf p'}}}\left ({F_{1}({\bf k},{\bf p},{\bf p'})
\over\omega_{{\bf p'}+{\bf p}}-\omega_{\bf p'}-\bar{\xi}_{{\bf p}+{\bf k}}}+{F_{2}({\bf k},{\bf p},{\bf p'})\over
\omega_{\bf p'}-\omega_{{\bf p}+{\bf p'}}-\bar{\xi}_{{\bf p}+{\bf k}}}\right .\nonumber\\
&+&\left .{F_{3}({\bf k},{\bf p},{\bf p'})\over\omega_{\bf p'}+\omega_{{\bf p}+{\bf p'}}
-\bar{\xi}_{{\bf p}+{\bf k}}}-{F_{4}({\bf k},{\bf p},{\bf p'})\over\omega_{\bf p'}+\omega_{{\bf p}+{\bf p'}}
+\bar{\xi}_{{\bf p}+{\bf k}}}\right ).
\end{eqnarray}
\end{subequations}
\end{widetext}
In this case, the effective normal-state pseudogap parameter is obtained from Eq. (\ref{pseudogap}) as
$\bar{\Delta}_{\rm pg}=(1/N)\sum_{\bf k}\bar{\Delta}_{\rm pg}({\bf k})$.

\begin{figure}[h!]
\center\includegraphics[scale=0.335]{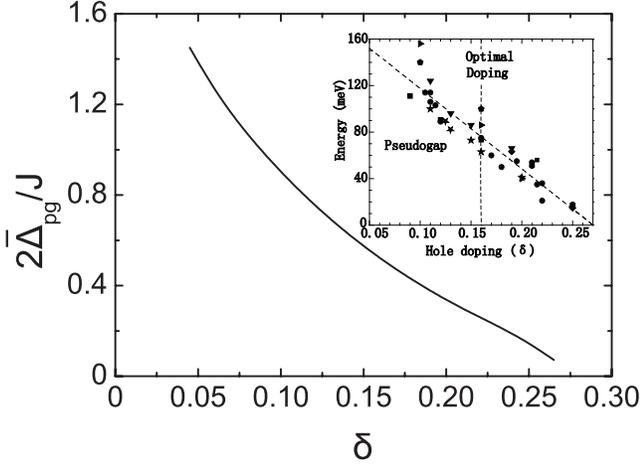}
\caption{The effective normal-state pseudogap parameter ($2\bar{\Delta}_{\rm pg}$) as a function of doping with
$T=0.002J$ for $t/J=2.5$ and $t'/t=0.3$. Inset: the corresponding experimental data of doped cuprates taken from Ref.
\onlinecite{Hufner08}. \label{fig1}}
\end{figure}

\begin{figure}[h!]
\center\includegraphics[scale=0.43]{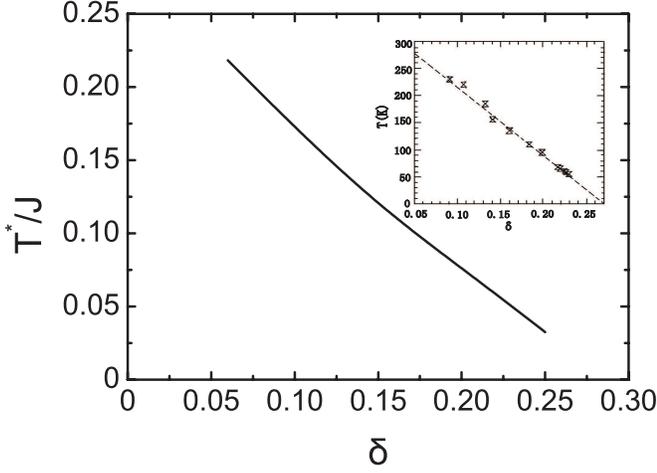}
\caption{The normal-state pseudogap crossover temperature $T^{*}$ as a function of doping for $t/J=2.5$ and $t'/t=0.3$.
Inset: the corresponding experimental data of Bi$_{2}$Sr$_{2}$CaCuO$_{8+\delta}$ taken from Ref. \onlinecite{Chatterjeea11}.
\label{fig2}}
\end{figure}

In doped cuprates, although the values of $J$, $t$, and $t'$ are believed to vary somewhat from compound to compound,
however, as a qualitative discussion in this paper, the commonly used parameters are chosen as $t/J=2.5$ and
$t'/t=0.3$. For a convenience in the following discussions of the charge dynamics, the calculated result of
$\bar{\Delta}_{\rm pg}$ versus doping for $T=0.002J$ is plotted in Fig. \ref{fig1} in comparison with the corresponding
experimental data \cite{Hufner08} observed on different families of doped cuprates (inset), where the magnitude of the
effective normal-state pseudogap parameter $\bar{\Delta}_{\rm pg}$ is particular large in the underdoped regime, then
it smoothly decreases upon increasing doping, in qualitative agreement with the experimental results \cite{Hufner08}.
Furthermore, this normal-state pseudogap is also temperature dependent. In particular, in the given doping
concentration, the normal-state pseudogap vanishes when temperature reaches the normal-state pseudogap crossover
temperature $T^{*}$. This $T^{*}$ satisfies the equation $0=(1/2N)\sum_{\bf k}L^{\rm (n)}_{2}({\bf k},T^{*})
/\sqrt{L^{\rm (n)}_{1}({\bf k},T^{*})}|_{\bar{\Delta}_{\rm pg}({\bf k},T^{*})=0}$, then $T^{*}$ and all other order
parameters are determined simultaneously by the self-consistent calculation \cite{feng12}. To analyze the evolution of
$T^{*}$ with doping, we have made a series of calculations for $T^{*}$ at different doping concentrations, and the
result of $T^{*}$ as a function of doping is plotted in Fig. \ref{fig2} in comparison with the corresponding
experimental data \cite{Chatterjeea11} observed on Bi$_{2}$Sr$_{2}$CaCuO$_{8+\delta}$ (inset). Our present result shows
that $T^{*}$ is much high in the underdoped regime, then it decreases with increasing doping, also in qualitative
agreement with the experimental results \cite{Chatterjeea11}. The essential physics of the present normal-state
pseudogap state is the same as the previous discussions of the two-gap feature \cite{feng12}, and can be attributed to
the doping and temperature dependence of the charge carrier interaction directly from the kinetic energy by exchanging
spin excitations.

Now we turn to discuss the charge dynamics in doped cuprates within the above microscopic theory of the normal-state
pseudogap state. The optical conductivity in the system is expressed as \cite{Mahan81},
\begin{eqnarray}\label{conductivity-1}
\sigma(\omega)=-{{\rm Im}\Pi(\omega)\over\omega},
\end{eqnarray}
where the electron current-current correlation function,
\begin{eqnarray}\label{correlation}
\Pi(\tau-\tau')=-\langle T_{\tau}{\bf j}(\tau)\cdot {\bf j}(\tau')\rangle,
\end{eqnarray}
with the electron current operator ${\bf j}$. The external magnetic field can be coupled to the electrons, which are
now represented by $C_{l\uparrow}=h^{\dagger}_{l\uparrow}S^{-}_{l}$ and
$C_{l\downarrow}= h^{\dagger}_{l\downarrow}S^{+}_{l}$ in the CSS fermion-spin representation. In this case, the
electron current operator is obtained in terms of the electron polarization operator, which is a summation over all
the particles and their positions, and is given in the CSS fermion-spin representation as ${\bf P}=-e\sum
\limits_{l\sigma}{\bf R}_{l}C^{\dagger}_{l\sigma}C_{l\sigma}=e\sum\limits_{l}{\bf R}_{l}h^{\dagger}_{l}h_{l}$, then
the electron current operator is obtained by evaluating the time-derivative of this polarization operator
\cite{Mahan81} ${\bf j}={\partial{\bf P}/\partial t}=(i/\hbar)[H,{\bf P}]$, and is evaluated explicitly as,
\begin{eqnarray}\label{tcurpara1}
{\bf j}&=&{iet\over\hbar}\sum\limits_{l\hat{\eta}}\hat{\eta}(h_{l\uparrow}h^{\dagger}_{l+\hat{\eta}\uparrow}
S^{+}_{l}S^{-}_{l+\hat{\eta}}+h_{l\downarrow}h^{\dagger}_{l+\hat{\eta}\downarrow}S^{-}_{l}S^{+}_{l+\hat{\eta}})
\nonumber\\
&-&{iet'\over\hbar}\sum\limits_{l\hat{\tau}}\hat{\tau}(h_{l\uparrow}h^{\dagger}_{l+\hat{\tau}\uparrow}S^{+}_{l}
S^{-}_{l+\hat{\tau}}+h_{l\downarrow}h^{\dagger}_{l+\hat{\tau}\downarrow}S^{-}_{l}S^{+}_{l+\hat{\tau}}).~~~~~~~
\end{eqnarray}
In the CSS fermion-spin approach, the electron current operator in Eq. (\ref{tcurpara1}) can be decoupled as,
\begin{eqnarray}\label{tcurpara2}
{\bf j}&=&-{ie\chi_{1}t\over\hbar}\sum\limits_{l\hat{\eta}\sigma}\hat{\eta}h^{\dagger}_{l+\hat{\eta}\sigma}
h_{l\sigma}+{ie\chi_{2}t'\over\hbar}\sum\limits_{l\hat{\tau}\sigma}\hat{\tau}h^{\dagger}_{l+\hat{\tau}\sigma}
h_{l\sigma}\nonumber\\
&-&{ie\phi_{1}t\over\hbar}\sum\limits_{l\hat{\eta}}\hat{\eta}(S^{+}_{l}S^{-}_{l+\hat{\eta}}+S^{-}_{l}
S^{+}_{l+\hat{\eta}})\nonumber\\
&+&{ie\phi_{2}t'\over\hbar}\sum\limits_{l\hat{\tau}}\hat{\tau}(S^{+}_{l}S^{-}_{l+\hat{\tau}}
+S^{-}_{l}S^{+}_{l+\hat{\tau}}),
\end{eqnarray}
where the charge carrier particle-hole parameters
$\phi_{1}=\langle h^{\dagger}_{l\sigma} h_{l+\hat{\eta}\sigma}\rangle$ and
$\phi_{2}=\langle h^{\dagger}_{l\sigma} h_{l+\hat{\tau}\sigma}\rangle$, while the third and fourth terms in the
right-hand side of Eq. (\ref{tcurpara2}) refer to the contribution from the electron spin, and are expressed
explicitly as,
\begin{eqnarray*}
&-&{ie\phi_{1}t\over\hbar}\sum\limits_{l,\hat{\nu}=\hat{x},\hat{y}}\hat{\nu}[(S^{+}_{l}S^{-}_{l+\hat{\nu}}
+S^{-}_{l}S^{+}_{l+\hat{\nu}})\\
&-&(S^{+}_{l}S^{-}_{l-\hat{\nu}}+S^{-}_{l}S^{+}_{l-\hat{\nu}})]\\
&=&-{ie\phi_{1}t\over\hbar}\sum\limits_{l,\hat{\nu}=\hat{x},\hat{y}}\hat{\nu}[(S^{+}_{l}S^{-}_{l+\hat{\nu}}
+S^{-}_{l}S^{+}_{l+\hat{\nu}})\\
&-&(S^{+}_{l+\hat{\nu}}S^{-}_{l}+S^{-}_{l+\hat{\nu}}S^{+}_{l})]\equiv 0,\\
&~&{ie\phi_{2}t'\over\hbar}\sum\limits_{l}[(\hat{x}+\hat{y})(S^{+}_{l}S^{-}_{l+\hat{x}+\hat{y}}+S^{-}_{l}
S^{+}_{l+\hat{x}+\hat{y}})\\
&-&(\hat{x}+\hat{y})(S^{+}_{l}S^{-}_{l-\hat{x}-\hat{y}}+S^{-}_{l}S^{+}_{l-\hat{x}-\hat{y}})\\
&+&(\hat{x}-\hat{y})(S^{+}_{l}S^{-}_{l+\hat{x}-\hat{y}}+S^{-}_{l}S^{+}_{l+\hat{x}-\hat{y}})\\
&-&(\hat{x}-\hat{y})(S^{+}_{l}S^{-}_{l-\hat{x}+\hat{y}}+S^{-}_{l}S^{+}_{l-\hat{x}+\hat{y}})]\\
&=&{ie\phi_{2}t'\over\hbar}\sum\limits_{l}[(\hat{x}+\hat{y})(S^{+}_{l}S^{-}_{l+\hat{x}+\hat{y}}+S^{-}_{l}
S^{+}_{l+\hat{x}+\hat{y}})\\
&-&(\hat{x}+\hat{y})(S^{+}_{l+\hat{x}+\hat{y}}S^{-}_{l}+S^{-}_{l+\hat{x}+\hat{y}}S^{+}_{l})\\
&+&(\hat{x}-\hat{y})(S^{+}_{l}S^{-}_{l+\hat{x}-\hat{y}}+S^{-}_{l}S^{+}_{l+\hat{x}-\hat{y}})\\
&-&(\hat{x}-\hat{y})
(S^{+}_{l+\hat{x}-\hat{y}}S^{-}_{l}+S^{-}_{l+\hat{x}-\hat{y}}S^{+}_{l})]\equiv 0,
\end{eqnarray*}
which reflects that within the framework of the CSS fermion-spin theory, the majority contribution for the electron
current operator comes from the charge carriers (then the electron charge), however, the strong interplay between the
charge carriers and spins has been considered through the spin's order parameters entering in the charge carrier part
of the contribution to the current-current correlation. In this case, the electron current-current correlation function
is evaluated in terms of the full charge carrier Green's function (\ref{Green's-function-1}) as,
\begin{eqnarray}\label{correlation-1}
\Pi(i\omega_{n})&=&-{1\over 2}(Ze)^{2}{1\over N}\sum_{\bf k}\gamma^{2}_{{\rm s}{\bf k}}\nonumber\\
&\times&{1\over\beta}\sum_{i\omega_{n'}}g({\bf k},i\omega_{n'}+i\omega_{n})g({\bf k},i\omega_{n'}),
\end{eqnarray}
where the current vertex $\gamma^{2}_{{\rm s}{\bf k}}=(1/4)[(\chi_{1}t-2\chi_{2}t'\cos k_{y})^{2}\sin^{2}k_{x}
+(\chi_{1}t-2\chi_{2}t'\cos k_{x})^{2}\sin^{2}k_{y}]$, and then the optical conductivity in Eq. (\ref{conductivity-1})
is obtained explicitly as,
\begin{eqnarray}\label{conductivity}
\sigma (\omega)&=&\left ({Ze\over 2}\right )^{2}{1\over N}\sum_{\bf k}\gamma_{{\rm s}{\bf k}}^{2}
\int^{\infty}_{-\infty}{{\rm d}\omega'\over 2\pi}A_{\rm h}({\bf k},\omega'+\omega)\nonumber\\
&\times&A_{\rm h}({\bf k},\omega')
{n_{\rm F}(\omega'+\omega)-n_{\rm F}(\omega')\over\omega},
\end{eqnarray}
with the charge carrier spectral function $A_{\rm h}({\bf k},\omega)=-2{\rm Im}g({\bf k},\omega)$.

\section{Doping and temperature dependence of conductivity} \label{transport}

\begin{figure}[h!]
\center\includegraphics[scale=0.34]{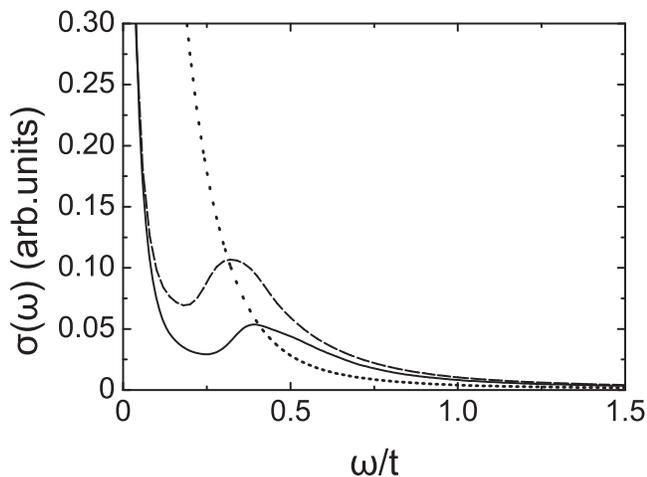}
\caption{The optical conductivity as a function of energy in $\delta=0.09$ (solid line), $\delta=0.15$ (dashed line),
and $\delta=0.25$ (dotted line) with $T=0.002J$ for $t/J=2.5$ and $t'/t=0.3$.\label{fig3}}
\end{figure}

\begin{figure}[h!]
\center\includegraphics[scale=0.34]{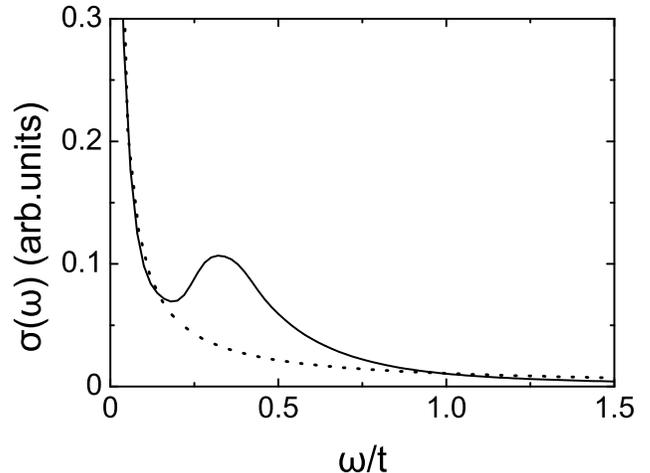}
\caption{The optical conductivity as a function of energy in $\delta=0.15$ with $T=0.002J$ for $t/J=2.5$ and
$t'/t=0.3$. The dashed line is obtained from a numerical fit $\sigma(\omega)=A/\omega$, with $A\sim 0.01$.
\label{fig4}}
\end{figure}

\begin{figure}[h!]
\center\includegraphics[scale=0.34]{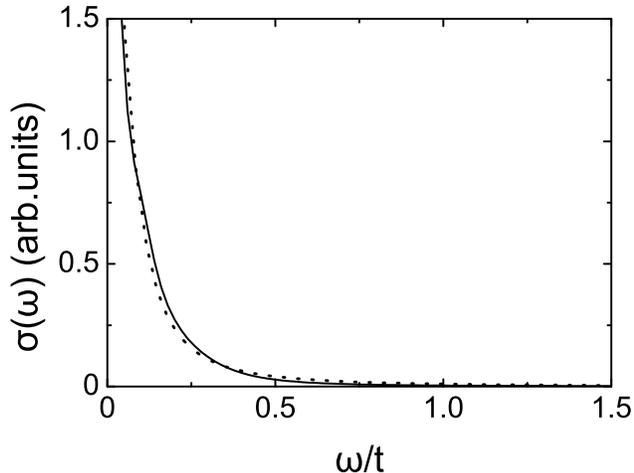}
\caption{The optical conductivity as a function of energy in $\delta=0.25$ with $T=0.002J$ for $t/J=2.5$ and
$t'/t=0.3$. The dashed line is obtained from a numerical fit $\sigma(\omega)=A/(\omega^{2}+B)$, with $A\sim 0.01$ and
$B\sim 0.0045$. \label{fig5}}
\end{figure}

We are now ready to discuss the doping and temperature dependence of the charge dynamics in doped cuprates. We have
performed a numerical calculation for the optical conductivity (\ref{conductivity}), and the results of
$\sigma(\omega)$ as a function of energy in the underdoping $\delta=0.09$ (solid line), the optimal doping
$\delta=0.15$ (dashed line), and the heavy overdoping $\delta=0.25$ (dotted line) with temperature $T=0.002J$
are plotted in Fig. \ref{fig3}, where the charge $e$ has been set as the unit. It is shown clearly that our present
theoretical results capture all qualitative features of the doping dependence of the optical conductivity
$\sigma(\omega)$ observed experimentally on doped cuprates
\cite{Tannern92,Cooper94,Kastner98,Timusk99,Puchkov96,Orenstein90,Uchida91,Puchkov96a,Basov96,Homes04}. In the
underdoped regime, there are two bands in $\sigma(\omega)$ separated by a gap at $\omega\sim 0.2t$, the higher energy
band, corresponding to the "midinfrared band", shows a broad peak at $\omega\sim 0.38t$. Moreover, the transferred
weight of the low-energy band forms a sharp peak at $\omega\sim 0$, which can be described formally by the non-Drude
formula. However, the weight and position of the midinfrared band are strongly doping dependent. In the optimal doping,
although two band features still are apparent, the positions of the gap and midinfrared peak appreciably shift towards
to the lower energies at $\omega\sim 0.16t$ and $\omega\sim 0.3t$, respectively, reflecting a tendency that with
increasing doping, the magnitude of the gap decreases, while the midinfrared band moves towards to the low-energy
non-Drude band. However, as in the case in the underdoped regime, the low-energy peak in the optimal doping still shows
the non-Drude formula. To see this point clearly, we have fitted our present result of $\sigma(\omega)$ in the optimal
doping $\delta=0.15$, and the result is shown in Fig. \ref{fig4}, where we found that the lower-energy peak decay as
$\rightarrow 1/\omega$. On the other hand, the tendency of the decrease of the magnitude of the gap and the midinfrared
band moving towards to the low-energy non-Drude band with increasing doping is particularly obvious in the overdoped
regime, in particular, the low-energy non-Drude peak incorporates with the midinfrared band in the heavily overdoped
regime, and then the low-energy Drude type behavior of the optical conductivity recovers, which is shown clearly in
Fig. \ref{fig5}, where we have also fitted the result of $\sigma(\omega)$ in the heavily overdoping $\delta=0.25$, and
the result indicates that in contrast with the case in the underdoped and optimally doped regimes, the lower-energy
peak decay as $\rightarrow 1/\omega^{2}$ in the heavily overdoped regime.

\begin{figure}[h!]
\center\includegraphics[scale=0.34]{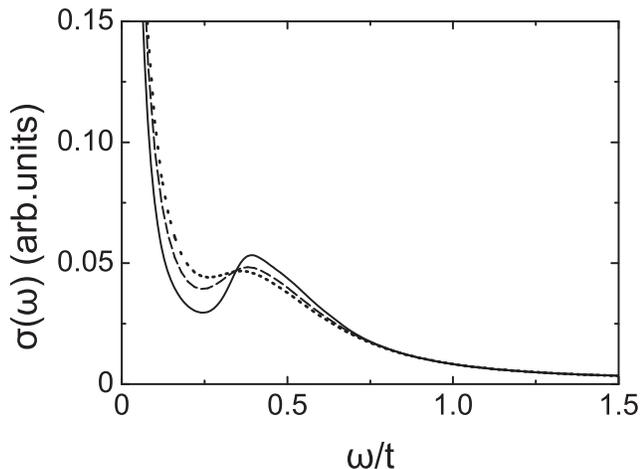}
\caption{The optical conductivity as a function of energy in $\delta=0.09$ with $T=0.02J$ (solid line), $T=0.146J$
(dashed line), and $T=0.186J$ (dotted line) for $t/J=2.5$ and $t'/t=0.3$.\label{fig6}}
\end{figure}

The low-energy non-Drude peak and unusual midinfrared band in the conductivity spectrum of doped cuprates in the
underdoped and optimally doped regimes is also temperature dependence. For a better understanding of the evolution of
the optical conductivity with temperature, we have further performed a calculation for $\sigma(\omega)$ in
Eq. (\ref{conductivity}) with different temperatures, and the results of $\sigma (\omega)$ as a function of energy
with $T=0.02J$ (solid line), $T=0.146J$ (dashed line), and $T=0.186J$ (dotted line) for $\delta=0.09$ are plotted in
Fig. \ref{fig6}. Within the present theoretical framework, the calculated normal-state pseudogap crossover temperature
$T^{*}\sim 0.19J$ at doping $\delta=0.09$. Our results show that the weight of the midinfrared band is severely
suppressed with increasing temperatures, and vanishes above the temperatures $T>T^{*}$, which are also qualitatively
consistent with the experimental data observed in doped cuprates
\cite{Tannern92,Cooper94,Kastner98,Timusk99,Puchkov96,Orenstein90,Uchida91,Puchkov96a,Basov96,Homes04}.

In an ordinary metal, the shape of the optical conductivity $\sigma (\omega)$ is normally well accounted for by the
low-energy Drude formula that describes the free charge carrier contribution to $\sigma (\omega)$. However, as shown
in Fig. \ref{fig1}, the part of the low-energy spectral weight in $\sigma (\omega)$ in the underdoped and optimally
doped regimes is transferred to the higher energy region to form the unusual midinfrared band due to the strongly
correlated nature in doped cuprates, then the width of the low-energy band is narrowing, while the onset of the region
to which the spectral weight is transferred, is close to the effective normal-state pseudogap $\bar{\Delta}_{\rm pg}$.
Since the unusual midinfrared band is taken from the low-energy band, so that both the low-energy non-Drude peak and
unusual midinfrared band describe the actual charge carrier density. In the CSS fermion-spin theory, the basic
low-energy excitations are the gauge invariant charge carrier and spin \cite{feng04}. However, the present results show
that main contribution to the charge transport in doped cuprates comes from the charge carriers, which are strongly
renormalized because of the interaction between the charge carriers and spins directly from the kinetic energy by
exchanging spin excitations. The $1/\omega$ decay of the optical conductivity at low energies in the underdoped and
optimally doped regimes is closely related with the linear temperature resistivity, since it reflects an anomalous
frequency dependent scattering rate proportional to $\omega$ instead of $\omega^{2}$ as would be expected in the
conventional Fermi-liquid.

The essential physics of the low-energy non-Drude peak and unusual midinfrared band in doped cuprates can be attributed
to the emergence of the normal-state pseudogap. As we have mentioned above, one quasiparticle band in the full charge
carrier Green's function in the absence of the normal-state pseudogap has been split into two branches due to the
presence of the normal-state pseudogap. In this case, the low-energy Drude peak in the conductivity spectrum in the
absence of the normal-state pseudogap is separated as the low-energy non-Drude peak and unusual midinfrared band due to
the quasiparticle band split in the presence of the normal-state pseudogap. However, the magnitude of the energy
difference between two subbands $\Delta({\bf k})=E^{+}_{{\rm h}{\bf k}}-E^{-}_{{\rm h}{\bf k}}$ in the full charge
carrier Green's function (\ref{Green's-function-3}) follows the same doping and temperature dependent behavior of the
normal-state pseudogap in Fig. \ref{fig1} and Fig. \ref{fig2}, i.e., it also decreases with increasing doping and
temperatures. In particular, the large energy difference in the underdoped regime leads to a strong separation between
the low-energy non-Drude peak and unusual midinfrared band. However, with increasing doping, the magnitude of the
normal-state pseudogap decreases as shown in Fig. \ref{fig1}, this leads to a decrease of the magnitude of the energy
difference between two subbands, and then the midinfrared band moves towards to the low-energy non-Drude band. In
particular, in the heavily overdoped regime, the normal-state pseudogap is very small, and therefore can be negligible,
which leads to that the energy difference between two subbands vanishes, and then the full charge carrier Green's
function (\ref{Green's-function-3}) is reduced approximately as,
\begin{eqnarray}\label{metal-form}
g({\bf k},\omega)&\approx&{1\over\omega-\xi_{{\bf k}}}.
\end{eqnarray}
In this case, the normal-state of doped cuprates is a conventional Fermi liquid similar to that of an ordinary metal,
then the unusual midinfrared band disappears, and the low-energy Drude type optical behavior recovers. This is also why
the unconventional charge transport appeared obviously in doped cuprates in the underdoped and optimally doped regimes
is absent in the heavily overdoped regime.

\section{Conclusions}\label{conclusions}

Within the microscopic theory of the normal-state pseudogap state, we have provide a natural explanation to the unusual
conductivity spectrum in doped cuprates. The conductivity spectrum in the underdoped and optimally doped regimes
contains the low-energy non-Drude peak and unusual midinfrared band. However, the position of the midinfrared band
shifts towards to the low-energy non-Drude peak with increasing doping. In particular, the low-energy non-Drude peak
incorporates with the midinfrared band in the heavily overdoped regime, and then the low-energy Drude behavior
recovers. The qualitative reproduction of all main features of the optical measurements on doped cuprates based on the
microscopic theory of the normal-state pseudogap state shows that the striking behavior of the low-energy non-Drude
peak and unusual midinfrared band in the underdoped and optimally doped regimes is closely related to the emergence of
the doping and temperature dependence of the normal-state pseudogap.

Finally, we have noted that within the framework of the preformed pair theory \cite{chen05}, the optical conductivity
in the underdoped cuprates has been discussed \cite{dan12}. In this preformed pair theory \cite{chen05}, the pair gap
contains both the condensed and noncondensed parts, with the SC-state is due to the condensation of the condensed pairs,
while the pseudogap state is associated with the part of the noncondensed pairs (then the preformed pairs), then the
transfer of the spectral weight from the low-energy peak to the midinfrared band in the underdoped cuprates can be
attributed to the emergence of this pseudogap \cite{dan12}. The origin of the pseudogap state in the preformed pair
theory \cite{chen05} is different from that suggested in our previous work \cite{feng12}, where the pseudogap state is
induced by the interaction between charge carriers and spins directly from the kinetic energy by exchanging spin
excitations in the particle-hole channel. However, In spite of the different origins of the pseudogap state between the
preformed pair theory \cite{chen05} and our previous work \cite{feng12}, the main feature of the charge carrier
propagators in both theories in the normal-state are very similar, this is why the results of the unusual conductivity
spectrum in the present work are qualitatively consistent with these obtained based on the the preformed pair theory
\cite{dan12}, then both theories indicate that the unusual conductivity spectrum is closely associated with the
pseudogap.

The doped cuprates have a layered structure consisting of the two-dimensional CuO$_{2}$ layers separated by insulating
layers. In this case, the effect of the vertex corrections for the conductivity spectrum is important, since the
self-energies in the two-dimensional system are strongly momentum dependent. However, based on the two-particle
self-consistent approach, the optical conductivity in the two-dimensional Hubbard model in the pseudogap regime has
been studied by considering the effect of the vertex corrections \cite{Bergeron11}, and the results show that although
the vertex corrections are important at all dopings, the typical hump structure due to the transfer of the spectral
weight from the low-energy peak to the midinfrared band in the midinfrared band range, related to the pseudogap, is
observed both with and without vertex corrections, but only with different amplitude at a given temperature. This
reflects that even the vertex corrections are dropped, the qualitative behavior of the hump structure of the
conductivity spectrum in the midinfrared band range due to the presence of the pseudogap is kept. In other words, the
hump structure in the midinfrared band range is mainly dominated by the pseudogap. On the other hand, we in this paper
are primarily interested in exploring the general notion of the effect of the pseudogap on the optical conductivity
based on the microscopic theory of the pseudogap state. The qualitative agreement between the present theoretical
results and experimental data also confirm that the origin of the transfer of the spectral weight from the low-energy
peak to the midinfrared band in doped cuprates is due to the presence of the pseudogap.

\acknowledgments

The authors would like to thank Dr. Huaisong Zhao for the helpful discussions. LQ and SF are supported by the National
Natural Science Foundation of China (NSFC) under Grant Nos. 11074023 and 11274044, and the funds from the Ministry of
Science and Technology of China under Grant Nos. 2011CB921700 and 2012CB821403, and JQ is supported by NSFC under Grant
No. 11004006.

\end{document}